\documentclass[12pt]{article}
\usepackage{amsmath,amssymb,bm,epsfig,color}
\renewcommand{\thefootnote}{\fnsymbol{footnote}}

\begin{document}

\title{
\begin{flushright}
\begin{minipage}{0.2\linewidth}
\normalsize
EPHOU-17-006 \\
WU-HEP-17-08 \\*[50pt]
\end{minipage}
\end{flushright}
{\Large \bf 
Supersymmetry preserving and  breaking  degenerate vacua, 
and radiative moduli stabilization
\\*[20pt]}}

\author{Tatsuo~Kobayashi,$^{1,}$\footnote{
E-mail address: kobayashi@particle.sci.hokudai.ac.jp}\ \
Naoya Omoto,$^{1,}$\footnote{
E-mail address: omoto@particle.sci.hokudai.ac.jp}\ \
Hajime~Otsuka,$^{2,}$\footnote{
E-mail address: h.otsuka@aoni.waseda.jp}\\
\ and \
Takuya~H.~Tatsuishi$^{1}$\footnote{
E-mail address: t-h-tatsuishi@particle.sci.hokudai.ac.jp}
\\*[20pt]
$^1${\it \normalsize 
Department of Physics, Hokkaido University, Sapporo 060-0810, Japan}  \\
$^2${\it \normalsize 
Department of Physics, Waseda University, 
Tokyo 169-8555, Japan
} \\*[50pt]}

\date{
\centerline{\small \bf Abstract}
\begin{minipage}{0.9\linewidth}
\medskip 
\medskip 
\small
We propose a new type of moduli stabilization scenario where the supersymmetric 
and supersymmetry-breaking minima are degenerate at the leading level. 
The inclusion of the loop-corrections originating from the matter fields resolves 
this degeneracy of vacua.  
Light axions are predicted in one of our models. 
%Axions may remain light in one of our models.
\end{minipage}
}

\begin{titlepage}
\maketitle
%\thispagestyle{empty}
%\clearpage
%\tableofcontents
%\thispagestyle{empty}
\end{titlepage}

\renewcommand{\thefootnote}{\arabic{footnote}}
\setcounter{footnote}{0}
%\vspace{35pt}

\section{Introduction}
The superstring theory predicts six-dimensional (6D) compact space in addition to 
four-dimensional (4D) spacetime.
The size and shape of the 6D compact space are determined by moduli.
Thus, moduli are a characteristic feature in the superstring theory on compact space.
Indeed, moduli  fields play important roles in the superstring theory and its 
4D low-energy effective field theory, in particular 
in particle phenomenology and cosmology.(See for a review, e.g. Refs.~\cite{Ibanez:2012zz,Blumenhagen:2006ci}.)
Studies on physics relevant to moduli would provide a remnant of the superstring theory on 
compact space.

Gauge couplings, Yukawa couplings and other couplings in the 4D low-energy effective field theory 
are given by vacuum expectation values (VEVs)  of moduli fields.
In the phenomenological point of view, the spectrum of supersymmetric particles 
is sensitive to the supersymmetry (SUSY) breaking of moduli fields through the gravitational 
interactions between the moduli fields and matter fields 
\cite{Kaplunovsky:1993rd,Brignole:1993dj,Choi:2005ge,Choi:2005hd}. 
This aspect would be relevant to dark matter physics, since 
the lightest superpartner is a candidate of dark matter.
Moreover, the thermal history of the Universe highly depends on the dynamics of 
moduli fields as well as axion fields, which are imaginary parts of moduli fields. 
From the theoretical point of view, 
these moduli fields are originating from the vector and tensor fields in 
low-energy effective action of the superstring theory as well as the higher-dimensional supergravity. 
The stabilization of the moduli field is one of the most important issues 
to realize a consistent low-energy effective action of the superstring theory.

The moduli potential is prohibited by the higher-dimensional 
gauge and Lorentz symmetries at the perturbative level.
On the other hand, the nontrivial background fields and nonperturbative effects 
generate the moduli potential. 
Then, one can stabilize the moduli fields and also study the dynamics relevant to moduli fields. 
The vacuum structure of the moduli potential is of particular importance.
For example, the flat direction of the moduli potential can drive cosmological inflation\footnote{See for the detail of moduli inflations as well as axion inflations, e.g., Ref.~\cite{Baumann:2014nda}.}
and the lifetime of our Universe depends on the (meta)stability of the vacuum. 

So far, there are several mechanisms to stabilize the moduli, in particular closed 
string moduli fields, e.g., the Kachru-Kallosh-Linde-Trivedi (KKLT) scenario \cite{Kachru:2003aw} 
and the LARGE volume scenario \cite{Balasubramanian:2005zx}. 
(See, e.g., Ref.~\cite{Ibanez:2012zz} and references therein.)
In this paper, we propose a new type of moduli stabilization scenario by using the 
string-derived ${\cal N}=1$ four-dimensional supergravity action. 
We find that the supersymmetric and SUSY-breaking vacua are degenerate at the tree level 
and they are independent of the $F$ term of certain moduli fields. 
The loop effects originating from the matter fields 
generate the moduli potential and resolve this degeneracy of vacua.

This paper is organized as follows.
In section~\ref{sec:2}, we study a new type of potential for a complex structure modulus 
within the framework of the type IIB superstring theory.
Its minimum has the SUSY preserving and breaking degenerate vacua, which can be 
resolved by loop effects.
Then, the modulus field including the axion is stabilized.
In section~\ref{sec:3}, we study similar aspects for the K\"ahler modulus. 
In this case, the real part of the K\"ahler modulus is stabilized, but the axion remains 
massless at this stage.
Finally, Sec.~\ref{sec:con} is devoted to the conclusion. 
In the Appendix, we give an illustrating model for the moduli stabilization.

\section{Complex structure moduli}
\label{sec:2}
In sections~\ref{sec:2} and~\ref{sec:3}, we consider two illustrative supergravity models where 
the moduli fields correspond to the complex structure modulus and K\"ahler modulus within 
the framework of the type IIB superstring theory. In both scenarios, we show that the supersymmetric and 
SUSY-breaking vacua are degenerate at the leading level. 
The inclusion of the one-loop corrections from the matter fields resolves this degeneracy.

In the type IIB superstring theory on the Calabi-Yau (CY) orientifold, 
the K\"ahler potential of moduli fields is 
described by
\begin{align}
K&=-\ln \left( i\int_{\rm CY} \Omega \wedge \bar{\Omega}\right)
-\ln (S+\bar{S}) -2\ln {\cal V},
%\nonumber\\
%W&=\int_{\rm CY} G_3 \wedge \Omega,
\end{align}
where $\Omega(U_m)$ is the holomorphic three-form of the CY manifold and 
${\cal V}(T_i)$ is the volume of the CY manifold.
% and $G_3=F_3-iSH_3$ is a 
%imaginary self-dual three-form. 
Here, $S$, $U_m$ and $T_i$ denote the dilaton, complex structure moduli
and K\"ahler moduli, respectively.

The three-form flux can induce the superpotential \cite{Gukov:1999ya}
\begin{align}\label{eq:W-flux}
%K&=-\ln \left( i\int_{\rm CY} \Omega \wedge \bar{\Omega}\right)
%-\ln (S+\bar{S}) -2\ln {\cal V},
%\nonumber\\
W_{\rm flux}&=\int_{\rm CY} G_3 \wedge \Omega,
\end{align}
where $G_3=F_3-iSH_3$ is an imaginary self-dual three-form. 
Also nonperturbative effects such as D-brane instantons and gaugino condensations 
can generate the superpotential of $S$ and $T_i$, e.g., 
\begin{equation}
W_{\rm np} = \sum_p A^{(p)}(U_m)e^{-a^{(p)}S - a^{(p)}_i T_i},
\label{eq:Wnp} 
\end{equation}
where $A^{(p)}(U_m)$ represent the $U_m$-dependent one-loop corrections and 
$a^{(p)}$ and $a^{(p)}_i$ are the numerical constants.

\subsection{The degenerate scalar potential}
\label{sec:2_1}
First, let us consider the following K\"ahler potential and superpotential 
based on the four-dimensional ${\cal N}=1$ supergravity:
\begin{equation}
K = -3 \ln (U +  \bar U), \qquad 
W = C_0 + C_1 U,
\label{eq:KW1}
\end{equation}
where $C_{0,1}$ are the complex constants. Throughout this paper, we use the reduced 
Planck unit $M_{\rm Pl}=2.4\times 10^{18}\,{\rm GeV}=1$. 
In the type IIB superstring theory on the CY orientifold, the modulus field $U$ 
could be identified with one of the complex structure moduli of the CY manifold. 
%Although the K\"ahler potential generically depends on the other complex structure moduli, 
We now assume that the other complex structure moduli and dilaton are stabilized at the 
minimum by the three-form fluxes~\cite{Giddings:2001yu}. 
When the parameters $C_{0,1}$ are determined only by the three-form fluxes as well as 
VEVs of the other complex structure moduli, these would be of ${\cal O}(1)$.
On the other hand, when the above superpotential appears from the instanton effects, 
$C_{0,1}$ are characterized as $e^{-aT}$, with $T$ being a certain K\"ahler modulus 
of the CY manifold, and could be suppressed. 
In the following analysis, we treat $C_{0,1}$ as complex constants by further assuming 
that all the K\"ahler moduli are stabilized at the minimum by the other nonperturbative effects. 
%Moreover, we assume that the dilaton $S$ is stabilized by the three-form flux or non-perturbative effects.
When $C_{0,1}$ are of the order of unity, it is a challenging issue to stabilize the 
K\"ahler modulus at the scale above the mass of $U$. Conversely, when both $C_{0,1}$ are exponentially suppressed, 
one can  achieve the above assumption as shown later.
In the following, we use the parametrization as $C_1=w_0$ and $C_0/C_1=C$. 
%\textcolor{red}{In both scenarios, we use the parametrization as $C_1=w_0$ and $C_0/C_1=C$.}

To see the degenerate supersymmetric and SUSY-breaking vacua, 
we calculate the scalar potential in the notation of $U=U_R+iU_I$ and  $C=C_R+iC_I$, 
\begin{equation}
V =e^K\left(K^{U\bar{U}}|D_UW|^2-3|W|^2\right)= - \frac{|w_0|^2}{6 U_R^2} (3C_R + 2 U_R),
\label{eq:VF}
\end{equation}
where $K^{U\bar{U}}$ is the inverse of K\"ahler metric $K_{U\bar{U}}=\partial_U\partial_{\bar{U}}K$ 
and 
\begin{equation}
D_UW = w_0\left(  -3 \frac{C + U}{U + \bar U} +1 \right).
\end{equation}
As a result, the scalar potential~(\ref{eq:VF}) remains flat in the direction of $U_I$. 
On the other hand, from the extremal condition of $U_R$,
\begin{equation}
\frac{\partial V}{\partial U_R} = \frac{|w_0|^2}{6 U_R^3} (6C_R + 2 U_R) =0,
\end{equation}
$U_R$ is stabilized at the minimum
\begin{equation}
U_{R,{\rm min}} = -3C_R,
\end{equation}
where $C_R$ should be negative to justify our low-energy effective action. 
{Note that $|C_R|$ is typically of the order of unity in both scenarios: $C_{0,1}\simeq {\cal O}(1)$ and $C_{0,1}\simeq {\cal O}(e^{-aT})$. 
If $|C_R|$ is smaller than unity, 
the moduli space of $U$ deviates from the large complex structure regime and our discussing logarithmic 
the K\"ahler potential is not reliable.} 
It turns out that the mass squared of canonically normalized $U_R$, 
\begin{equation}
\frac{K^{U\bar{U}}}{2}\frac{\partial^2 V}{\partial U_R^2} = \frac{2 U_{R,{\rm min}}^2}{3}\frac{|w_0|^2}{3 U_{R,{\rm min}}^3}
=\frac{2|w_0|^2}{9 U_{R,{\rm min}}},
\end{equation}
is positive and the vacuum energy, 
\begin{equation}
V = -\frac{|w_0|^2}{6 U_{R,{\rm min}}},
\end{equation}
is negative at this minimum. 
Note that the mass squared of canonically normalized $U_R$ at this vacuum is taken smaller 
than the other moduli fields, in particular, the K\"ahler moduli. 
For example, when we consider the nonperturbative superpotential for the K\"ahler moduli irrelevant to our focusing $U$, the K\"ahler moduli can be stabilized 
at certain minima by them and the mass squared of the overall K\"ahler modulus is given by ${\cal O}((2\pi {\rm Re}(T))^2 |w_0|^2$) for the KKLT scenario. It is much heavier than that of our considered $U_R$ as 
discussed in the Appendix~\ref{app}.

Although the vacuum energy is independent of $U_I$, 
the $F$-term of the modulus $U$ depends on $U_I$, 
\begin{equation}
F=-e^{K/2}K^{U\bar{U}}D_{\bar{U}}\bar{W},
\end{equation}
where 
\begin{equation}
{\rm Re}(D_UW) = 0, \qquad {\rm Im}(D_U W) = w_0 \left(-\frac{3 (C_I + U_I)}{2U_{R,{\rm min}}}\right).
\end{equation}
Thus, we find that the SUSY is preserved at $C_I + U_I =0$ 
and broken at $C_I + U_I \neq 0$, respectively. 
From the fact that the scalar potential is independent of $U_I$, 
supersymmetric and SUSY-breaking vacua are degenerate. 
%Such a degeneracy is possible at the anti de Sitter vacuum.

\subsection{Loop corrections}
\label{sec:2_2}
The analysis in Sec.~\ref{sec:2_1} shows that the vacuum energy is negative, $V<0$, 
and at the same time, the scalar potential remains flat in the direction of $U_I$. 
First, we introduce the uplifting sector to obtain the tiny cosmological constant. 
In particular, we assume that the $U$-independent potential 
induced by the anti-D-branes uplifts the anti-de Sitter vacuum to the Minkowski one such as the KKLT scenario.\footnote{Uplifting by spontaneous $F$-term SUSY breaking is also possible \cite{Lebedev:2006qq,Dudas:2006gr}.
Even in that case, the $U_I$ direction would remain flat.} 
Next, the nonvanishing $F$ term of $U$ gives rise to the soft terms of matter fields. 
These massive supersymmetric particles induce the $U_I$-dependent scalar potential 
through one-loop corrections \cite{Coleman:1973jx}:
\begin{equation}
\Delta V = {\rm Str }\frac{{\cal M}^4}{64\pi^2 } \ln \left[ {\cal M}^2/ \Lambda^2 \right]. 
\end{equation}

In the following analysis, we illustrate how we can stabilize $U_I$ by 
one-loop effects. 
For such a purpose, we focus on the situation that the gauginos and 
supersymmetric scalar fields contribute to the one-loop corrections. 
The gaugino masses are provided by
\begin{equation}
M_a = \frac{1}{2} \frac{\partial \ln f_a}{\partial \ln \Phi^I} F^I ,
\end{equation}
where $f_a(\Phi^I)$ with $a=U(1)_Y, SU(2)_L, SU(3)_C$ 
representing the gauge kinetic functions 
for the standard model gauge groups, respectively. 
%We may include gaugino masses in the hidden sector.
On the other hand, the soft scalar masses are given by
\begin{equation}
m^2_i = \frac23 V_0- {\partial_I}{\partial_{\bar I}}Y_{i\bar i} |F^I|^2 + ({\rm D\ term}),
\end{equation}
where
\begin{equation}
Y_{i \bar i} = e^{-K/3}Z_{i \bar i},
\end{equation}
with $Z_{i \bar i}$ being the K\"ahler metric of the matter fields. 

To simplify our illustrating analysis, we assume that the typical soft scalar mass and gaugino mass 
mainly contribute to the one-loop potential. 
Then, those are characterized as
\begin{align}
M= k_f F, \qquad m^2 =  m^2_0 - k_m |F|^2,
\label{eq:soft}
\end{align}
where $k_f$ and $k_m$ are real constants and $m^2_0$ denotes the soft scalar mass 
induced by the $U$-independent $F$-term contributions. 
The gaugino mass may also have another contribution such as $M = k_fF + M_0$.
Even in such a case, the following discussion is similar.
For simple illustration, we restrict ourselves to the above spectrum of superpartners.

\begin{figure}[htbp]
  \begin{center}
    \begin{tabular}{c}

      % potential form
      \begin{minipage}{0.5\hsize}
        \begin{center}
          \includegraphics[clip, width=7.0cm]{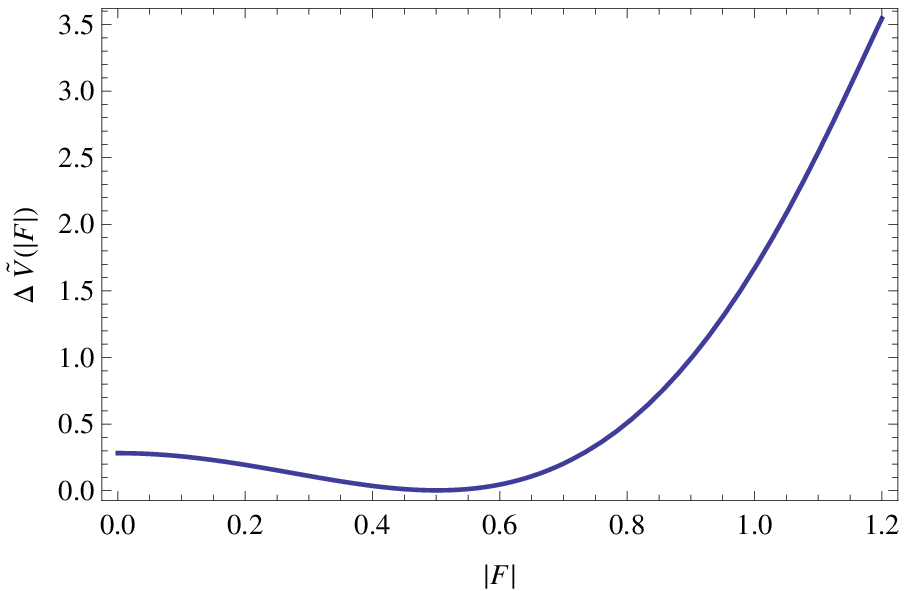}
          \hspace{1.6cm} 
        \end{center}
      \end{minipage}

     % trajectory
      \begin{minipage}{0.5\hsize}
        \begin{center}
          \includegraphics[clip, width=7.0cm]{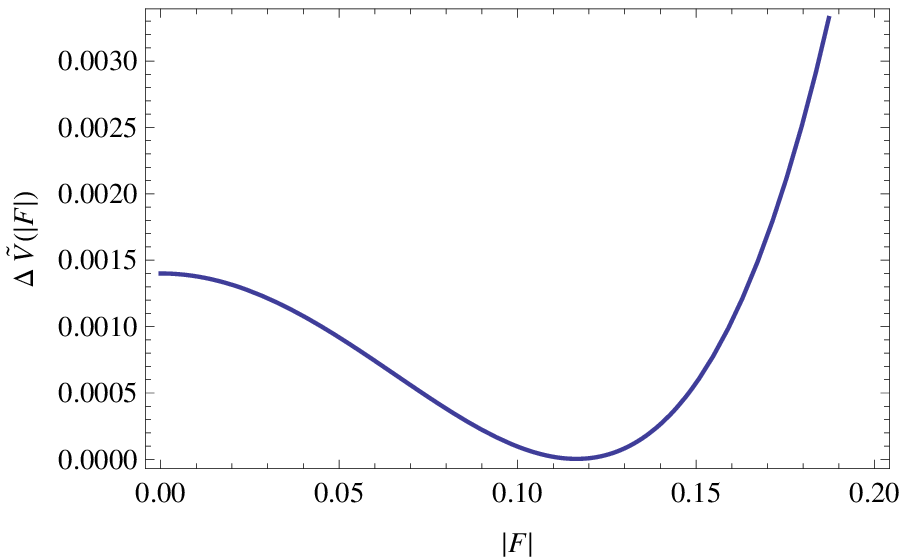}
          \hspace{1.6cm}
        \end{center}
      \end{minipage}

    \end{tabular}
    \caption{The one-loop scalar potential as a function of $|F|$. 
    The parameters are set as $a_0=1,a_3=0.1,c=1.2$ in the left panel and 
    $a_0=-1,a_3=0.1,c=0.2$ in the right panel.}
    \label{fig:potential2}
%     \ref{fig: potential2}
  \end{center}
\end{figure}

\begin{figure}[htbp]
  \begin{center}

      \begin{minipage}{0.8\hsize}
        \begin{center}
          \includegraphics[clip, width=7.0cm]{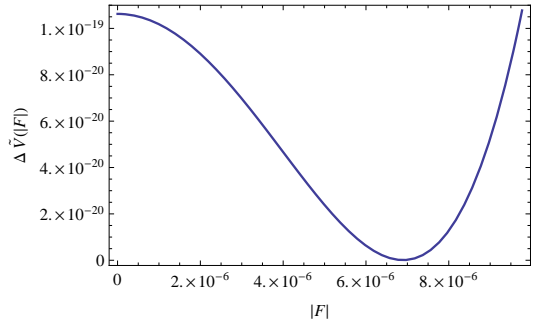}
          \hspace{1.6cm} 
        \end{center}
      \end{minipage}

  \caption{The one-loop scalar potential as a function of $|F|$ in the case of small $c$. 
 The parameters are taken as $a_0=-1,a_3=0.1,c=10^{-5}$.}
    \label{fig:potential3}
    % \ref{fig: potential3}
  \end{center}
\end{figure}

By rescaling the $F$ term of $U$, one can set $k_m=1$, 
and the corresponding one-loop potential can be written by 
\begin{equation}
64\pi^2 \Delta V=a_1(c^2-|F|^2)^2\ln (c^2-|F|^2)-a_2F^4\ln (a_3|F|^2) +V_0,
\end{equation}
where $c^2=m^2_0/k_m$, $a_3=k_f/k_m$ and $a_{1,2}$ correspond to the multiplicities of the scalars and 
gauginos, respectively. 
Now, we include the constant $V_0$ coming from the $F$ terms of  K\"ahler moduli and anti-D-brane effects to achieve the tiny cosmological constant at the vacuum. 
By using $\Delta \tilde V = 64 \pi^2 \Delta V/a_2$, $a_0=a_1/a_2$, and $\tilde{V}_0=V_0/a_2$, the above scalar potential is simplified as
\begin{equation}
 \Delta \tilde V=a_0(c^2-|F|^2)^2\ln (c^2-|F|^2)-F^4\ln (a_3|F|^2) +\tilde{V}_0.
\end{equation}
The first derivative of the one-loop potential with respect to $|F|$ is 
given by
\begin{align}
\Delta \tilde{V}'&= -2|F|\biggl[ 2|F|^2\ln (a_3 |F|^2)+|F|^2+a_0(c^2-|F|^2)+2a_0(c^2-|F|^2)\ln(c^2-|F|^2)\biggl],
\label{eq:second deriv.2}
%\ref{eq:second deriv.2}
\end{align}
from which there are two possible minima leading to $|F|=0$ and $|F|\neq 0$. 
To see the nonvanishing $F$, we draw the one-loop scalar potential as a function of $|F|$ by setting the following illustrative parameters:
\begin{align}
a_0=1(-1),\qquad
a_3=0.1,\qquad
c=1.2(0.2),\qquad
\tilde{V}_0\simeq -0.474(-0.00375),
\end{align}
in the left (right) panel in Fig.~\ref{fig:potential2} and
\begin{align}
a_0=-1,\qquad
a_3=0.1,\qquad
c=10^{-5},\qquad
\tilde{V}_0\simeq -1.24\times 10^{-19},
\end{align}
in Fig.~\ref{fig:potential3}. 
It turns out that the nonvanishing of $|F|$ depends on the sign of $a_0$ and 
the value of $c$.

%%%%%%%%%%%%%%%%%%%%%%%%%%%%%%%%%%%%%%%%%%%%%%%%%%%%%%%%%%%%%%%%%%%%%%%%%%%%%%%%%%%%%%%%%%%%%%%%
%%%%%%%%%%%%%%%%%%%%%%%%%%%%%%%%%%%%%%%%%%%%%%%%%%%%%%%%%%%%%%%%%%%%%%%%%%%%%%%%%%%%%%%%%%%%%%%%
%%=========%
%%=========%

%=========%
%=========%
%=========%
%%%%%%%%%%%%%%%%%%%%%%%%%%%%%%%%%%%%%%%%%%%%%%%%%%%%%%%%%%%%%%%%%%%%%%%%%%%%%%%%%%%%%%%%%%%%%%%
%%%%%%%%%%%%%%%%%%%%%%%%%%%%%%%%%%%%%%%%%%%%%%%%%%%%%%%%%%%%%%%%%%%%%%%%%%%%%%%%%%%%%%%%%%%%%%%%
To discuss the vacuum structure of the one-loop scalar potential, 
we analytically derive the condition of vanishing $|F|$. 
From the second derivative of the one-loop potential with respect to $|F|$
\begin{align}
\Delta \tilde{V}''&= -2\left(a_0 c^2-7(a_0-1)|F|^2+6F^2\ln(a_3 |F|^2)+2a_0(c^2-3|F|^2)\ln(c^2-|F|^2)\right),
\end{align}
$\Delta \tilde{V}''$ becomes negative at the origin $|F|=0$ for 
 \begin{eqnarray}
%\left.\tilde{V}''\right|_{|F|=0} \ < \  0&\Longleftrightarrow& 
-2\left(a_0 c^2+2a_0c^2\ln c^2\right)\ < \  0.
\end{eqnarray}
It implies that when $a_0 >0$ and 
\begin{eqnarray}
c \ > \ \frac{1}{e^{1/4}}\simeq 0.78,
\end{eqnarray} 
the minimum of $|F|$ is taken as ${\cal O}(1)$ because of the instability at $|F|=0$. 
However, we assume that the other moduli fields are decoupled from our system 
and such a high-scale SUSY breaking is not reliable. 
Thus, when $a_0>0$, supersymmetric minimum $|F|=0$ is favorable. 
In this case, the soft terms are determined by $F$ terms of the K\"ahler moduli. 
Such a vanishing $F$ term of $U$ is also interesting from the aspects of 
the flavor structure of the matter fields. 
Indeed, Yukawa couplings among the standard model particles 
depend on the complex structure moduli through the compactification of 
an extra dimension as derived in the type IIB superstring theory with magnetized D-branes~\cite{Cremades:2004wa}. 
Thus, a sizable $F$ term of a complex structure modulus is dangerous for 
the flavor-changing processes among the supersymmetric particles, 
which are severely constrained in the low-scale SUSY-breaking scenario. 

%%%%%%%%%%%%%%%%%%%%%%%%%%%%%%%%%%%%%%%%%%%%%%%%%%%%%%%%%%%%%%%%%%%%%%%%%%%%%%%%%%%%%%%%%%%%%%%%%%%
Let us take a closer look at the vacuum structure of the one-loop scalar potential. 
First, we further simplify the scalar potential by setting 
\begin{align}
V_{\rm sim}\equiv \Delta \tilde{V}/c^4,\qquad
V_{\rm sim}^{(0)}\equiv \tilde{V}_0/c^4,\qquad
{\cal C}\equiv \ln c^2,\qquad
A_3\equiv a_3c^2,\qquad
\mathcal{F}\equiv |F|/c.
\end{align}
Then, the total scalar potential consist of the radiative corrections from 
bosonic and fermionic parts:
\begin{align}
V_{\rm sim}=V_b({\cal F})+V_f({\cal F})+V_{\rm sim}^{(0)},
\label{eq:Vsim1}
\end{align}
where 
\begin{align}
V_b({\cal F})&=a_0(1-{\cal F}^2)^2 \biggl[{\cal C}+\ln (1-{\cal F}^2)\biggl],
\nonumber\\
V_f({\cal F})&=-{\cal F}^4 \ln (A_3 {\cal F}^2).
\label{eq:VbVf}
\end{align}
Note that $|F|>c$ gives rise to the tachyonic soft scalar masses from the 
definition of soft scalar mass in Eq.~(\ref{eq:soft}) and $c=m_0^2/k_m$. 
Thus, ${\cal F}=|F|/c$ is constrained to the range, $0\leq {\cal F}<1$. 

\begin{figure}[h]
  \centering
   \includegraphics[width=70mm]{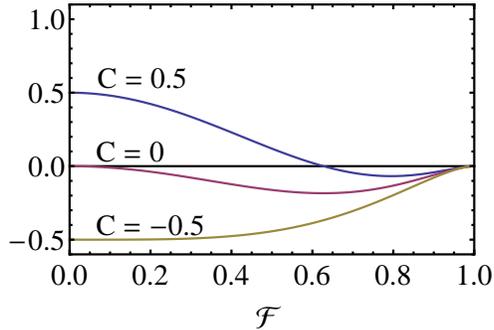}
\caption{{The bosonic part $V_b$}.}
\label{fig:boson}
\end{figure}

%%%%%%%%%%%%%%%%%%%%%%%%%%%%%%%%%%%%%%%%%%%%%%%%%%%%%%%%%%%%%%%%%%%%%%%%%%%%%%%%%%%%%%%%%%%%%%%%%%%
First of all, let us discuss the bosonic contribution. 
From the first derivative of $V_b({\cal F})$ with respect to ${\cal F}$,
\begin{align}
\frac{\partial V_b({\cal F})}{\partial{\cal F}}&=-4a_0{\cal F}(1-{\cal F}^2)
\biggl[\ln (1-{\cal F}^2)+{\cal C}+\frac{1}{2}\biggl],
\end{align}
we find that
there are two possible SUSY and SUSY-breaking vacua in the physical domain $0\leq {\cal F}<1$. 
When ${\cal C}\leq -1/2$, the SUSY is preserved as
\begin{align}
{\cal F}=0,
\end{align}
whereas the SUSY is broken under ${\cal C}>-1/2$, 
\begin{align}
{\cal F}=\sqrt{1-e^{-({\cal C}+1/2)}}. 
\end{align}
It can be confirmed in Fig.~\ref{fig:boson}, where 
we show the parameter dependence of the bosonic {part} $V_b$.  

\begin{figure}[h]
  \centering
   \includegraphics[width=100mm]{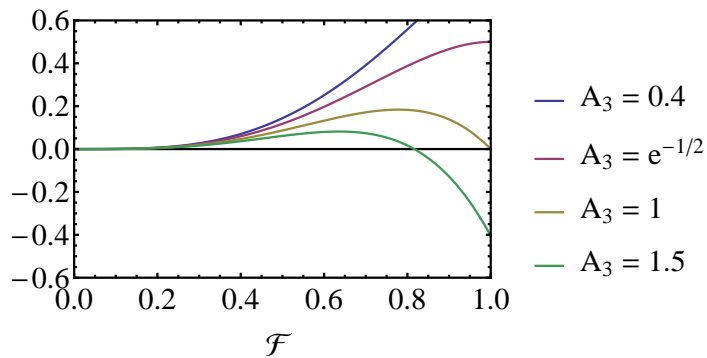}
\caption{{The fermionic part $V_f$.}}
\label{fig:fermion}
\end{figure}

%%%%%%%%%%%%%%%%%%%%%%%%%%%%%%%%%%%%%%%%%%%%%%%%%%%%%%%%%%%%%%%%%%%%%%%%%%%%%%%%%%%%%%%%%%%%%%%%%%
Next, we focus on the fermionic contribution. 
From the first derivative of $V_f({\cal F})$ with respect to ${\cal F}$,
\begin{align}
\frac{\partial V_f({\cal F})}{\partial{\cal F}}&=-2{\cal F}^3
\biggl[2\ln (A_3{\cal F}^2)+1\biggl],
\end{align}
we find that there are two minima 
\begin{align}
{\cal F}=0,
\end{align}
and
\begin{align}
{\cal F}=e^{-1/4}A_3^{-1/2}. 
\end{align}
However, the latter one leads to the tachyonic mass squared, 
$\partial_{\cal F}\partial_{\cal F} {\cal V}_f=-8(e^{-1/4}A_3^{-1/2})^2<0$. 
Thus the fermionic contribution leads to the SUSY-preserving minimum 
without depending on the value of $A_3$. 
In Fig.~\ref{fig:fermion}, the fermionic part $V_f$ is drawn as a function of $A_3$. 
It turns out that $A_3$ larger than $e^{-1/2}$ causes the instability of the vacuum. 

\begin{figure}[h]
 \begin{minipage}{0.5\hsize}
  \centering
   \includegraphics[width=80mm]{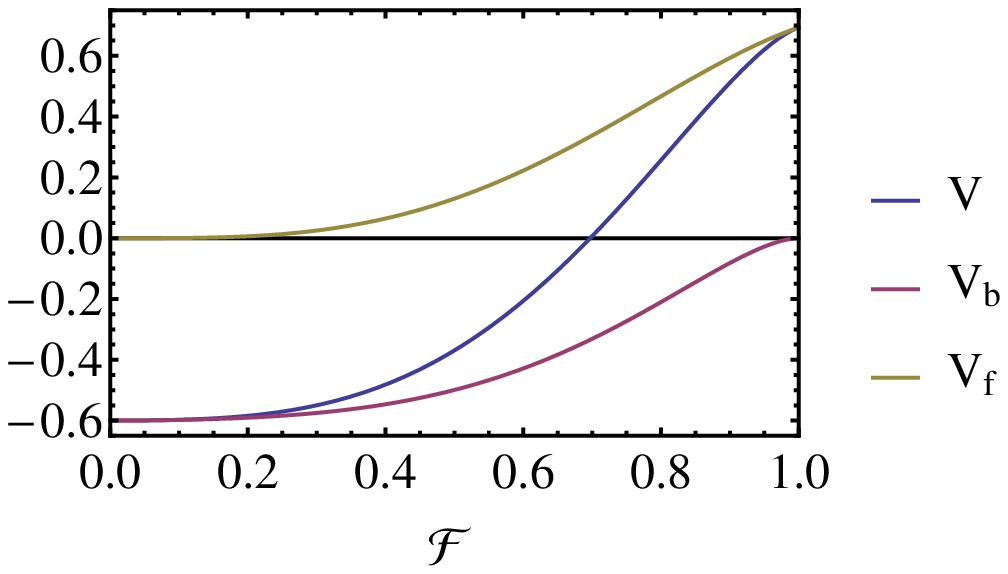}
 \end{minipage}
 \begin{minipage}{0.5\hsize}
  \centering
  \includegraphics[width=80mm]{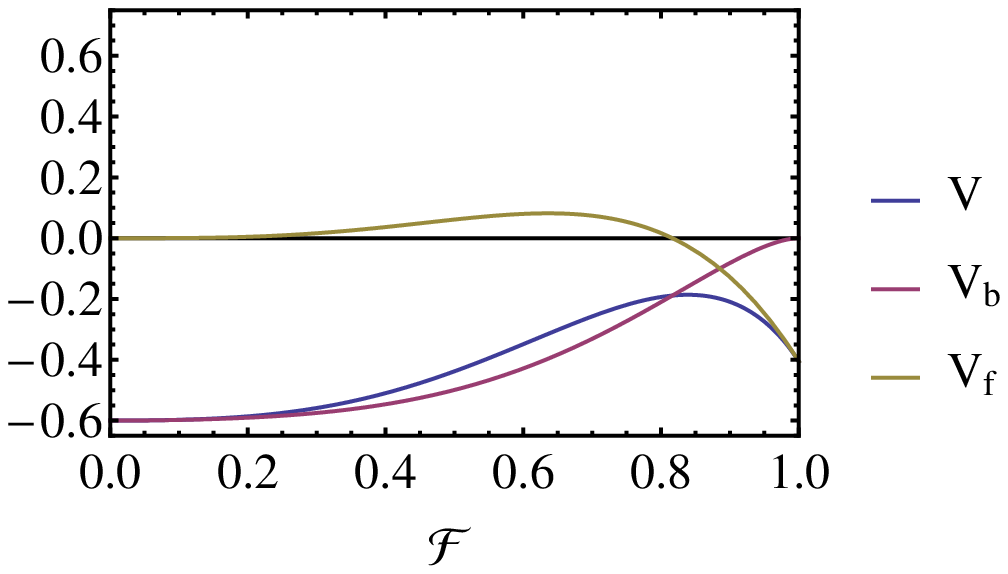}
 \end{minipage}
\caption{{The potential forms with $a_0=1,\,C=-0.6,\,A_3=0.5,\,(c=0.74,\,a_3=0.91)$ in the left panel and 
 $a_0=1,\,C=-0.6,\,A_3=1.5,\,(c=0.74,\,a_3=2.73)$ in the right panel.}}
\label{fig:bf1}
\end{figure}

%%%%%%%%%%%%%%%%%%%%%%%%%%%%%%%%%%%%%%%%%%%%%%%%%%%%%%%%%%%%%%%%%%%%%%%%%%%%%%%%%%%%%%%%%%%%%%%%%%
{Now, we combine } the bosonic and fermionic contributions. 
When ${\cal C}\leq -1/2$, the bosonic part is the monotonically increasing 
function with respect to ${\cal F}$, whereas the fermionic one depends on 
the magnitude of $A_3$. 
As shown in the left and right panels in Fig.~\ref{fig:bf1}, 
the vacuum instability arises by large values of $A_3$ included in the fermionic contribution. 
Since the vacuum energy is provided by
%\begin{align}
%V_1(0)=a_0 {\cal C}+V_0,
%\end{align}
%one can achieve the tiny cosmological constant by setting $V_0\simeq -a_0{\cal C}$. 

\begin{align}
V_{\rm sim}(0)=a_0 {\cal C}+V_{\rm sim}^{(0)},
\end{align}
one can achieve the tiny cosmological constant by setting $V_{\rm sim}^{(0)}\simeq -a_0{\cal C}$.

\begin{figure}[h]
 \begin{minipage}{0.5\hsize}
  \centering
   \includegraphics[width=80mm]{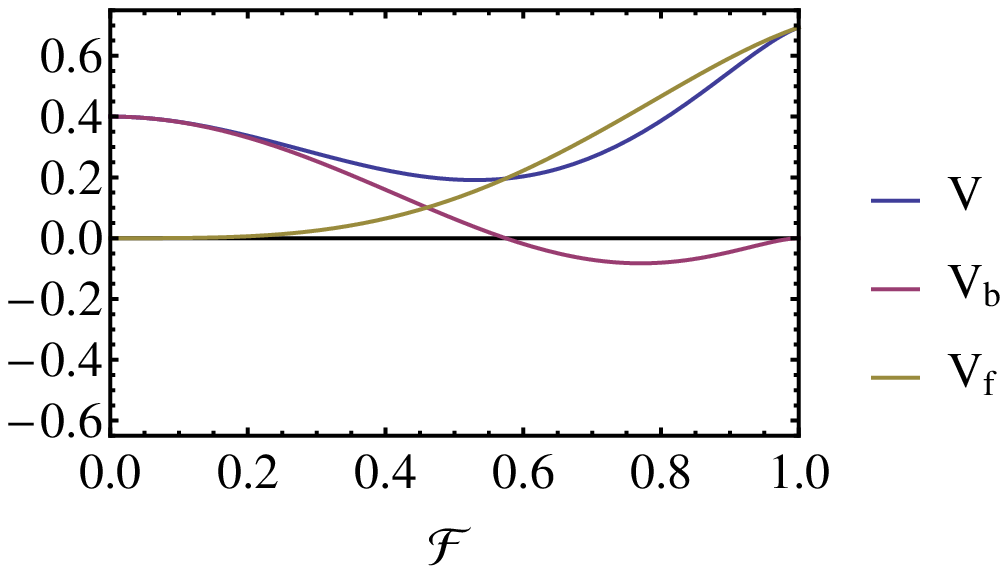}
 \end{minipage}
 \begin{minipage}{0.5\hsize}
  \centering
  \includegraphics[width=80mm]{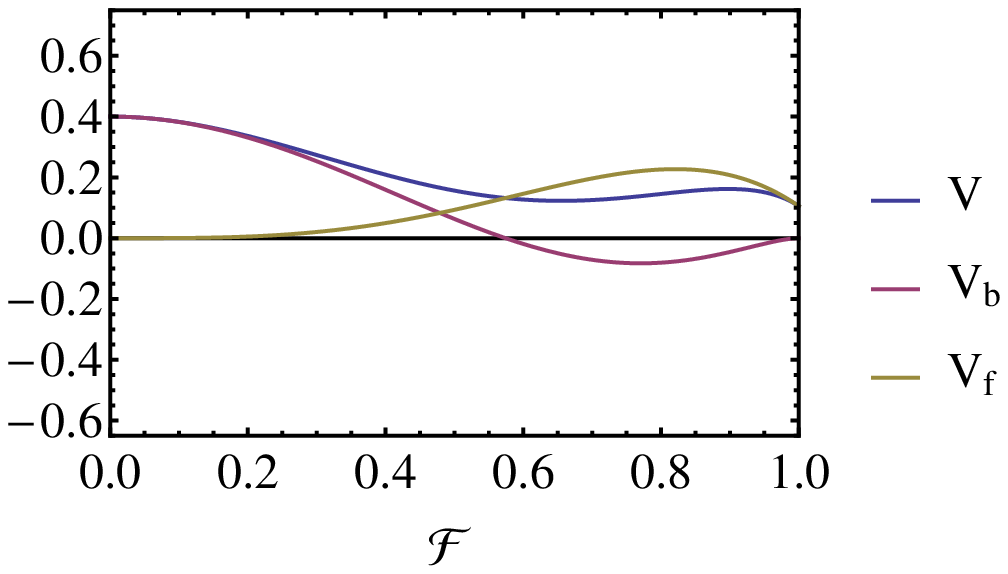}
 \end{minipage}
\caption{{The potential forms with $a_0=1,\,C=0.4,\,A_3=0.5,\,(c=1.22,\,a_3=0.34)$ in the left panel and 
 $a_0=1,\,C=0.4,\,A_3=0.9,\,(c=1.22,\,a_3=0.60)$ in the right panel.}}
  \label{fig:bf2}
\end{figure}

On the other hand, in the case of ${\cal C}> -1/2$, the vacuum structure is determined by $A_3$. 
Indeed, when $A_3<e^{-1/2}$, the vacuum of $V_b$ is shifted 
to the origin by the fermionic contribution as shown in the left panel in Fig.~\ref{fig:bf2}. 
However, the large value of $A_3>e^{-1/2}$ triggers the instability of the vacuum 
as can be seen in the right panel in Fig.~\ref{fig:bf2}. 

So far, we have discussed the case with $a_0>0$. 
It is found that the bosonic contribution gives the source of SUSY breaking, whereas 
the fermionic contribution preserves the SUSY irrelevant to the parameters. 
However, a large fermionic contribution causes the instability of the vacuum. 
In a similar way, we can discuss the case of $a_0<0$, which changes the bosonic part of the radiative corrections 
as $-V_b$.


%%%%%%%%%%%%%%%%%%%%%%%%%%%%%%%%%%%%%%%%%%%%%%%%%%%%%%%%%%%%%%%%%%%%%%%%%%%%%%%%%%%%%%%%%%%%%%%%%%%

\clearpage

\begin{figure}[htbp]
  \begin{center}
    \begin{tabular}{c}

      % potential form
      \begin{minipage}{0.5\hsize}
        \begin{center}
          \includegraphics[clip, width=7.0cm]{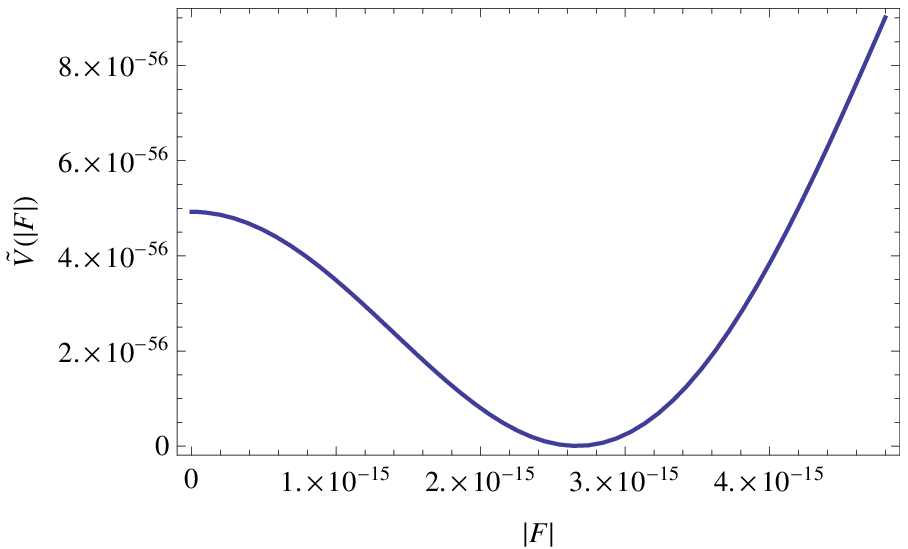}
          \hspace{1.6cm} 
        \end{center}
      \end{minipage}

     % trajectory
      \begin{minipage}{0.5\hsize}
        \begin{center}
          \includegraphics[clip, width=7.0cm]{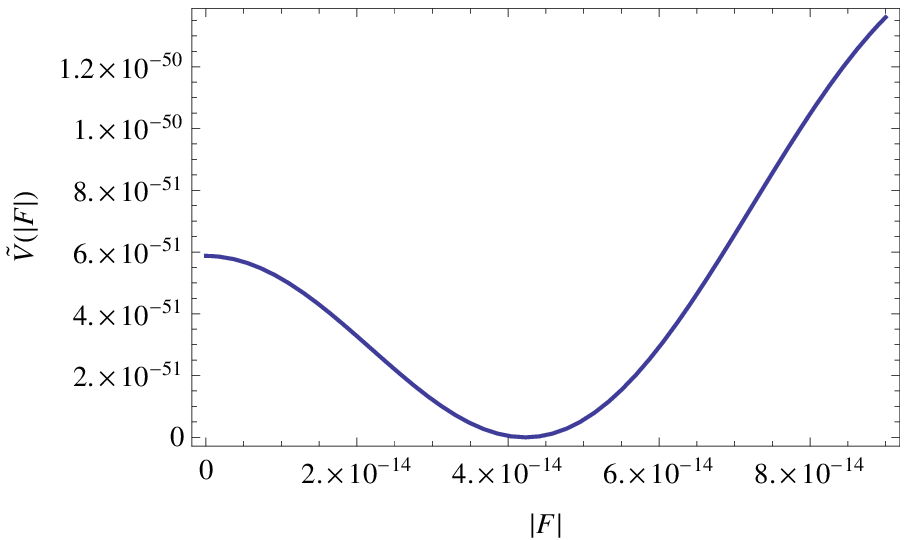}
          \hspace{1.6cm} 
        \end{center}
      \end{minipage}

    \end{tabular}
    \caption{The one-loop scalar potential involving the nonperturbative correction in Eq.~(\ref{eq:ponon}) 
    is drawn as a function of $|F|$. 
    The parameters are set as $a_0=a_3=1,a_4=10^{-15},c=5\times 10^{-15}, \tilde{\Lambda}=1.4\times 10^{-14}, \tilde{V}_1=5.2\times 10^{-56}$ in the left panel and 
    $a_0=a_3=1,a_4=10^{-15},c=9\times 10^{-14}, \tilde{\Lambda}=2.5\times 10^{-13}, \tilde{V}_1=5.9\times 10^{-51}$ in the right panel.}
    \label{fig:potential4}
    % \ref{fig: potential2}
  \end{center}
\end{figure}

Finally, we discuss the possibility of nonvanishing $F$ by adding the nonperturbative effects of $U$ 
into the one-loop scalar potential. 
The nonperturbative effects of $U$ are expected to appear through e.g. the gaugino condensation on 
hidden D-branes where the gauge kinetic function involves the $U$-dependent one-loop corrections~\cite{Lust:2003ky}. 
Since the real part of $U$ is already stabilized by the superpotential in Eq.~(\ref{eq:KW1}), 
the potential of $U_I$ can be extracted as
\begin{equation}
\Lambda^4 \cos (U_I/f + \theta_0),
\end{equation}
where $f$ is the typical decay constant and $\theta_0$ is a real constant. 
In the following, we set $C_I=0$ for simplicity. 
Then, the $F$ term of $U$,
\begin{equation}
|F^U| = e^{K/2} K^{U \bar U} |D_U W| = 9W_0U_I,
\end{equation}
leads to the following total scalar potential:
\begin{equation}
V=\frac{1}{64\pi^2}\biggl[a_1(c^2-|F|^2)^2\ln (c^2-|F|^2)-a_2|F|^4\ln (a_3|F|^2)\biggl]
+\Lambda^4 \cos \left(\frac{|F|}{9W_0f}\right)+V_1. 
\end{equation}
Here, the constant $V_1$ is inserted to realize the tiny cosmological constant at the vacuum 
in {a way} similar to the previous scenario. 
By rescaling the parameters as
\begin{align}
\tilde{V}\equiv 64\pi^2V/a_2,
\qquad
a_0\equiv a_1/a_2,
\qquad
\tilde{\Lambda}\equiv \Lambda (64\pi^2/a_2)^{1/4},
\qquad
a_4\equiv 9W_0f,
\qquad
\tilde{V}_1\equiv V_1 (64\pi^2/a_2),
\end{align}
we analyze the following potential
\begin{equation}
\tilde{V}=a_0(c^2-|F|^2)^2\ln (c^2-|F|^2)-F^4\ln (a_3|F|^2)
+\tilde{\Lambda}^4 \cos \left(\frac{|F|}{a_4}\right)+\tilde{V}_1. 
\label{eq:ponon}
\end{equation}
Figure.~\ref{fig:potential4} shows that the nonvanishing $|F|$ is achieved even when $a_0$ is positive. 
Since the origin of $c$ and $\tilde{\Lambda}$ are the nonperturbative effects, one can 
realize the low-scale SUSY-breaking scenario in this model.

%%%%%%%%%%%%%%%%%%%%%%%%%%%%%%%%%%%%%%%%%%%%%%%%%%%%%%%%%%%%%%%%%%%%%%%%%%%%%%%%%%%%%%%%%%%%%%%%
%%%%%%%%%%%%%%%%%%%%%%%%%%%%%%%%%%%%%%%%%%%%%%%%%%%%%%%%%%%%%%%%%%%%%%%%%%%%%%%%%%%%%%%%%%%%%%%%

We have assumed that loop corrections are dominant in the potential of $U_I$.
When other nonperturbative effects are dominant, obviously $U_I$ is stabilized by
such nonperturbative effects and loop effects provide subdominant corrections.

%%%%%%%%%%%%%%%%%%%%%%%%%%%%%%%%%%%%%%%%%%%%%%%%%%%%%%%%%%%%%%%%%%%%%%%%%%%%%%%%%%%%%%%%%%%%%%%%
%%%%%%%%%%%%%%%%%%%%%%%%%%%%%%%%%%%%%%%%%%%%%%%%%%%%%%%%%%%%%%%%%%%%%%%%%%%%%%%%%%%%%%%%%%%%%%%%
\section{K\"ahler moduli}
\label{sec:3}
In this section, we consider another example where the supersymmetric and 
SUSY-breaking minima are degenerate at the leading level. 

%In type IIB superstring theory on CY manifold, 
%K\"ahler potential and superpotential of moduli fields are 
%described by
%\begin{align}
%K&=-\ln \left( i\int_{\rm CY} \Omega \wedge \bar{\Omega}\right)
%-\ln (S+\bar{S}) -2\ln {\cal V},
%\nonumber\\
%W&=\int_{\rm CY} G_3 \wedge \Omega,
%\end{align}
%where $\Omega(U)$ is the holomorphic three-form of CY manifold, 
%${\cal V}(T)$ is the volume of CY manifold and $G_3=F_3-iSH_3$ is a 
%imaginary self-dual three-form. 
%Here, $S$, $U$ and $T$ denote the dilaton, complex structure moduli
%and K\"ahler moduli, respectively. 

By use of the flux-induced superpotential (\ref{eq:W-flux}), the $F$-term scalar potential is 
calculated as
\begin{align}
V_F&=e^K\biggl[\sum_{I,J=S,U_m}K^{I\bar{J}}D_IW D_{\bar{J}}\bar{W}+\left(K^{T_i\bar{T_j}}K_{T_i}K_{\bar{T_{j}}}-3\right)|W|^2
\biggl]
\nonumber\\
&=e^K\biggl[\sum_{I,J=S,U_m}K^{I\bar{J}}D_IW D_{\bar{J}}\bar{W}\biggl],
\end{align}
where $-3|W|^2$ is canceled by the no-scale structure of the K\"ahler moduli,
\begin{align}\label{eq:no-scale}
\sum_{i,j}K^{T_{i}\bar{T_j}}K_{T_i}K_{\bar{T_j}}-3=0.
\end{align}
Note that the above no-scale structure is valid only at the tree level. 

Then, the dilaton and complex structure moduli are stabilized 
at the minimum,
\begin{align}
D_SW=0,
\qquad
D_{U_m}W=0,
\end{align}
which lead to the Minkowski minimum $V_F=0$. 
When $W\neq 0$, the supersymmetry is broken by the $F$ term of the K\"ahler moduli. 
In contrast to the previous section, we now assume that all the complex structure moduli 
and dilaton are stabilized by the flux-induced superpotential. 
Although the $F$ terms of $S$ and $U$ vanish at this Minkowski minimum, 
the $F$ terms of the K\"ahler moduli are nonvanishing, in general:
\begin{align}
F^{T_i}=-e^{K/2}\sum_jK^{T_i\bar{T_j}}D_{\bar{T_j}}\bar{W}
=-e^{K/2}\sum_jK^{T_i\bar{T_j}}K_{\bar{T_j}}\bar{W},
\end{align}
when $W\neq 0$.

For simplicity, we study the model with the overall  K\"ahler modulus with 
the CY volume ${\cal V}=(T+\bar{T})^{3/2}$.
Then, the $F$ term of the K\"ahler modulus is simplified as
\begin{align}
F^T\simeq e^{K(S,U)/2}\frac{T+\bar{T}}{(T+\bar{T})^{3/2}}\bar{W}
=e^{K(S,U)/2}\frac{\bar{W}}{(T+\bar{T})^{1/2}}.
\end{align}
Thus, supersymmetric and SUSY-breaking minima are also degenerate in a way similar 
to the previous section, since the scalar potential is independent of $T$ and $F^T$. 
However, the supersymmetric vacuum corresponds to ${\rm Re}(T) \rightarrow \infty$, that is, 
the decompactification limit.

When the leading $\alpha^\prime$ corrections are involved, 
the K\"ahler potential of the K\"ahler modulus is corrected as~\cite{Becker:2002nn}
\begin{align}
K=-2\ln \left( {\cal V}+\frac{\xi}{2}\right),
\end{align}
where $\xi =-\frac{\chi (CY)\zeta(3)}{2(2\pi)^3g_s^{3/2}}$ with $\chi$ and 
$g_s$ being the Euler characteristic of CY and string coupling. 
These $\alpha^\prime$-corrections break the no-scale structure, 
and the scalar potential is generated as
\begin{align}
V_F&\simeq e^{K(S,U)}\frac{3\xi}{4{\cal V}^3}|W|^2.
\end{align}
The sign of $\xi$ depends on the number of complex structure moduli and 
K\"ahler moduli. When the number of K\"ahler moduli is smaller than that of 
complex structure moduli, $\xi$ is positive. 
In the case of single K\"ahler modulus, the $F$-term potential reduces to
\begin{align}
V_F\simeq e^{K(S,U)}\frac{3\xi}{4(T+\bar{T})^{9/2}}|W|^2=\frac{3\xi}{4(T+\bar{T})^{7/2}}|F^T|^2.
\end{align}

Along the same step outlined in Sec.~\ref{sec:2}, 
we take into account the loop corrections originating from the supersymmetric particles 
whose soft terms are dominated by the $F$ term of the K\"ahler modulus. 
It is remarkable that the loop corrections give rise to the stabilization of 
${\rm Re}(T)$ unlike the case in Sec.~\ref{sec:2}.
Then, by assuming that the typical gaugino and supersymmetric scalar fields mainly 
contribute to the loop effects, the total scalar potential becomes
\begin{align}
V&\simeq \frac{3\xi}{4(T+\bar{T})^{7/2}}|F^T|^2
+\frac{1}{64\pi^2}\biggl[a_1\left(c^2-\left(\frac{|F^T|}{T+\bar{T}}\right)^2\right)^2\ln \left(c^2-\left(\frac{|F^T|}{T+\bar{T}}\right)^2\right)
\nonumber\\
&-a_2\left(\frac{|F^T|}{T+\bar{T}}\right)^4\ln \left(a_3\left(\frac{|F^T|}{T+\bar{T}}\right)^2\right)\biggl]
\nonumber\\
&=  \frac{3\xi}{4(T+\bar{T})^{3/2}}(\hat{F}^T)^2
+\frac{1}{64\pi^2}\biggl[a_1\left(c^2-\left(\hat{F}^T\right)^2\right)^2\ln \left(c^2-\left(\hat{F}^T\right)^2\right)
\nonumber\\
&-a_2\left(\hat{F}^T\right)^4\ln \left(a_3\left(\hat{F}^T\right)^2\right)\biggl]
\nonumber\\
&=  \frac{3\xi}{4e^{K(S,U)/2}W}(\hat{F}^T)^3
+\frac{1}{64\pi^2}\biggl[a_1\left(c^2-\left(\hat{F}^T\right)^2\right)^2\ln \left(c^2-\left(\hat{F}^T\right)^2\right)
\nonumber\\
&-a_2\left(\hat{F}^T\right)^4\ln \left(a_3\left(\hat{F}^T\right)^2\right)\biggl],
\end{align}
where $\hat{F}^T\equiv |F^T|/(T+\bar{T})$. 
Here, we employ the same notation of Sec.~\ref{sec:2} and $W$ is chosen as a real constant, for simplicity.

\begin{figure}[htbp]
\centering
  \includegraphics[width=7.0cm]{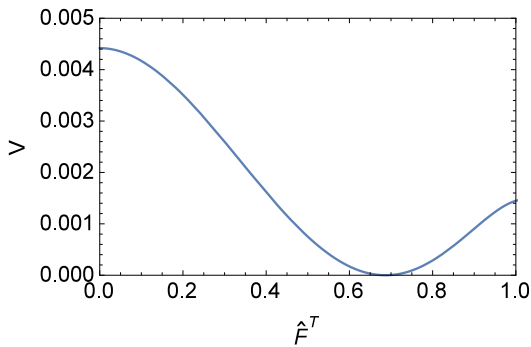}
    \caption{The scalar potential as a function of $\hat{F}^T$ by setting the parameters as in Eq.~(\ref{eq:paraT}).}
  \label{fig:T}
\end{figure}

By setting the illustrative parameters 
\begin{align}
a_1=10,
\qquad
a_2=3,
\qquad
a_3=8,
\qquad
c=1.1,
\qquad
\xi=1,
\qquad
e^{K(S,U)/2}W\simeq 60.42,
\label{eq:paraT}
\end{align}
the scalar potential is drawn as in Fig.~\ref{fig:T}. 
As a result, the degeneracy of  vacua is resolved by the loop corrections. 
In contrast to the discussion in Sec.~\ref{sec:2}, the vanishing $|F^T|\propto (T+\bar{T})^{-1/2}$ corresponds to 
the unphysical domain ${\rm Re}(T)\rightarrow \infty$. 
Thus, the SUSY-breaking vacuum is selected. 
Indeed, the above illustrative parameters give rise to the high-scale SUSY-breaking minimum, 
where the vacuum expectation value of ${\rm Re}(T)$,
\begin{align}
{\rm Re}(T)\simeq 9.9, 
\end{align}
resides in a reliable range of the supergravity approximation. 
After canonically normalizing the modulus,
\begin{align}
\hat{\sigma}=\sqrt{\frac{3}{2}}\ln \sigma, 
\end{align}
with $\sigma={\rm Re}(T)/\sqrt{2}$, its mass squared is evaluated as
\begin{align}
m^2_{\hat{\sigma}}\simeq 3.3\times 10^{-2},
\end{align}
in the reduced Planck unit, i.e. $m_{\hat{\sigma}}\simeq 0.18 \times M_{\rm Pl}$, which should be smaller than the other complex structure moduli and dilaton 
to justify our low-energy effective action. 
Since those complex structure moduli and dilaton fields have been stabilized at the SUSY-breaking 
minimum, their masses are typically greater than or equal to the gravitino mass $e^{K(S,U)/2}W/{\cal V}\simeq 6.9\times 10^{-1}$ in our numerical example. Our situation is thus justified.

Interestingly, the tuning of $e^{K(S,U)/2}W$ allows us to consider the tiny cosmological constant. 
In the above scenario, ${\rm Re}(T)$ can be stabilized at a fine value, but 
its imaginary part, i.e. the axion, remains massless.

%However, imaginary part of axion remains massless. 
%It is interesting to identify this massless axion with the candidate of inflaton, QCD axion and the candidate 
%of axion-inflaton. 
%The detailed discussion of axion phenomenology will be left for a future work. 

\begin{figure}[h]
 \begin{minipage}{0.5\hsize}
  \centering
   \includegraphics[width=80mm]{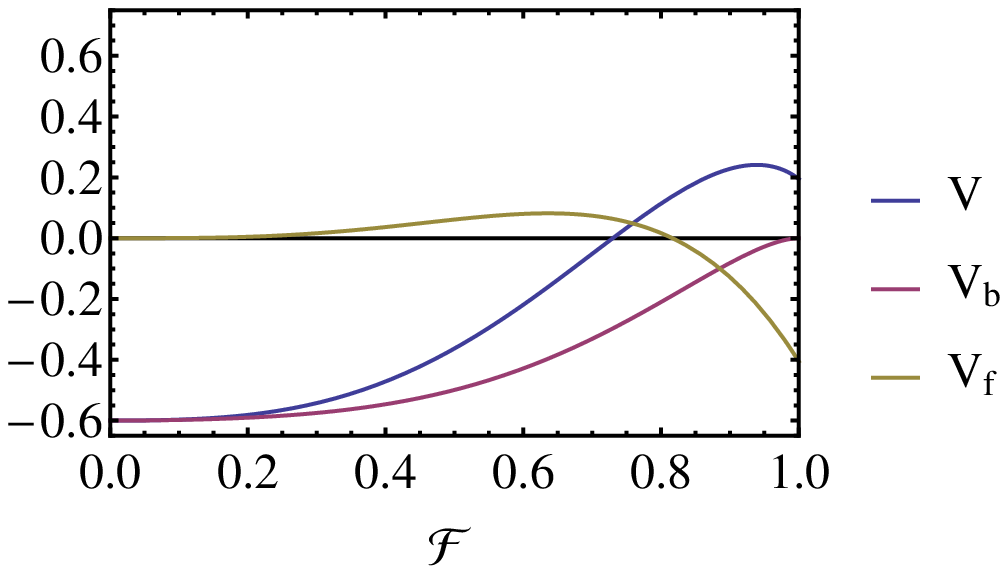}
 \end{minipage}
 \begin{minipage}{0.5\hsize}
  \centering
  \includegraphics[width=80mm]{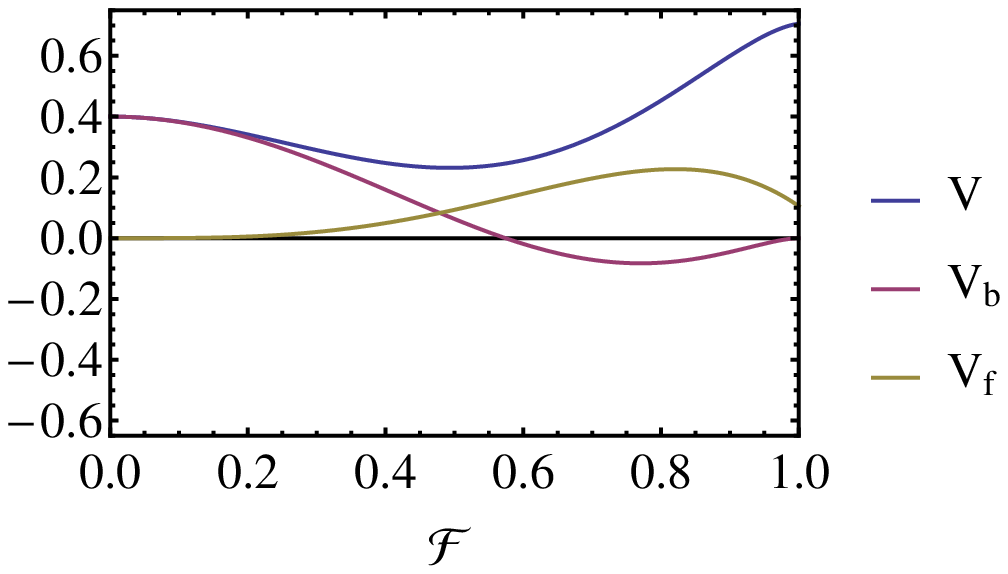}
 \end{minipage}
\caption{{The potential forms with $a_0=1,\,C=-0.6,\,A_3=1.5,\,A=0.6,\,(c=0.74,\,a_3=2.73,\,\alpha=1.67)$ 
in the left panel and $a_0=1,\,C=0.4,\,A_3=0.9,\,A=0.6,\,(c=1.22,\,a_3=0.60,\,\alpha=3.75)$ in the right panel.}}
\label{fig:bf3}
\end{figure}

Let us take a closer look at the scalar potential. 
To simplify our analysis, we redefine the scalar potential as a 
function of ${\cal F}\equiv \frac{\hat{F}^T}{c^2}$ which is constrained to be 
$0\leq {\cal F}<1$ to avoid the tachyonic soft scalar masses,
\begin{align}
V_{\rm sim}=64\pi^2 \frac{V}{a_2c^4}=V_b+V_f+A{\cal F}^3,
\end{align}
where
\begin{align}
V_b&\equiv a_0(1-{\cal F}^2)^2 \biggl[{\cal C}+\ln (1-{\cal F}^2)\biggl],
\nonumber\\
V_f&\equiv -{\cal F}^4 \ln (A_3 {\cal F}^2),
\end{align}
and the corresponding parameters are defined as
\begin{align}
a_0\equiv \frac{a_1}{a_2},
\qquad
{\cal C}\equiv \ln c^2,
\qquad
A_3\equiv a_3c^2,
\qquad
A\equiv \frac{(48\pi^2)\xi}{e^{K(S,U)/2}Wa_2}c^2.
\end{align}
The simplified scalar potential is similar to the scalar potential in Eq.~(\ref{eq:Vsim1}). 
However, in the current case, there exists a cubic term of ${\cal F}^3$ and the cosmological 
constant is not included. 
As outlined in Sec.~\ref{sec:2}, we concentrate on the case with $a_0>0$. 
When ${\cal C}\leq -1/2$, the scalar potential is drawn as in the left panel in Fig.~\ref{fig:bf3}, where 
we take the same parameters as in Fig.~\ref{fig:bf1}. 
The vacuum is settled into {${\cal F}=0$}. 
On the other hand, in the case of ${\cal C}> -1/2$, {the} cubic term of ${\cal F}$ lifts 
the potential compared with the case in Sec.~\ref{sec:2}. 
In the right panel in Fig.~\ref{fig:bf3}, we plot the scalar potential by taking the same parameters in Fig.~\ref{fig:bf2}. 
%Thus, the fermionic contribution gives the source of SUSY-breaking and at the same time 
%the instability of the vacuum. 

Thus, the bosonic contribution gives the source of SUSY breaking, and the fermionic contribution gives
the instability of the vacuum.

So far, we have studied the model with the overall K\"ahler modulus.
However, we can discuss the model with many K\"ahler moduli fields $T_i$.
Even in such a model, the tree-level scalar potential is flat 
along all of the K\"ahler moduli directions because of the no-scale structure (\ref{eq:no-scale}).
Also, the tree-level potential is independent of $F$ terms of $T_i$, 
and these $F^{T_i}$ themselves depend only on ${\rm Re}(T_i)$ but not ${\rm Im}(T_i)$.
In such multi-moduli models,  gaugino masses and soft scalar masses would be written by 
\begin{eqnarray}
M_a = M_0 + \sum_i k_f^i F^{T_i}, \qquad m_i^2 = m^2_0  - \sum_{i} k_m^i |F^{T_i}|^2. 
\end{eqnarray} 
Using these, we obtain the one-loop potential $\Delta V(T_i + \bar T_i)$.
Then, one can stabilize $F^{T_i}$ and ${\rm Re}(T_i)$ in a similar way.
However, all the axionic parts of $T_i$ remain massless at this stage. 

Let us consider the axion potentials by adding the nonperturbative effects such 
as the gaugino condensation on hidden D-branes or world-sheet instanton effects. 
Those nonperturbative effects appear in the superpotential as shown in Eq.~(\ref{eq:Wnp}), and 
the potentials of the canonically normalized axions $\psi^i$ are extracted as
\begin{eqnarray}
V=\sum_p D_p \cos(d_i^{(p)} \psi^i+\theta_p), 
\label{eq:axionpo}
\end{eqnarray} 
where $D_p$, $d_i^{(p)}$, and $\theta_p$ are the real constants determined by the stabilization of the complex 
structure moduli, dilaton, and the real parts of the K\"ahler moduli, respectively. 
If the axions appear in the above nonperturbative effects, their masses 
are much smaller than the other moduli fields.
Light axions would be interesting.
For example, they can be candidates for the QCD axion and dark matter. 
In such a case, the QCD axion appears only in the gauge kinetic functions of the visible sector and 
not in the nonperturbative effects~(\ref{eq:Wnp}). 
To identify the K\"ahler axion with the QCD axion and dark matter, the axion decay constant is relatively larger 
than the desirable one. For that reason, we require certain dilution effects below the QCD scale.
%If a proper potential for one of axions is generated below the above energy scale, 
%the axion may derive inflationary expansion of the Universe.
{If the axion potentials are generated as in Eq.~(\ref{eq:axionpo}), 
the axion may derive inflationary expansion of the Universe by using the alignment mechanism~\cite{Kim:2004rp}.
Furthermore, the string axiverse scenario may be realized if they are light \cite{Arvanitaki:2009fg}.\footnote{See for a realization of light axions, e.g., Refs.~\cite{Higaki:2011me,Halverson:2017deq}.}
%However, imaginary part of axion remains massless. 
%It is interesting to identify this massless axion with the candidate of inflaton, QCD axion and the candidate 
%of axion-inflaton. 
{At any rate, these aspects of axion phenomenology and cosmology highly depend on details of 
nonperturbative effects (\ref{eq:axionpo}) below the energy scale, where  the other moduli are stabilized.
Such an analysis is beyond our scope of this paper. }
The detailed discussion of axion phenomenology will be left for a future work.

\section{Conclusion}
\label{sec:con}
%In this paper, we have discussed the string inspired supergravity models 
%where the supersymmetric and SUSY-breaking mimima are degenerate 
%at the leading level. 
%The inclusion of the loop-corrections originating from the matter fields 
%resolves this degeneracy. 

We have studied a new type of moduli potential and stabilization.
In the model with the complex structure modulus $U$, 
 the supersymmetric and SUSY-breaking minima are degenerate at the leading order.
That is, the tree-level potential is independent of the $F$ term of $U$, but depends on ${\rm Re}(U)$.
The $F$ term itself depends on ${\rm Im}(U)$.
Loop effects due to ${\rm Im}(U)$-dependent gaugino and sfermion masses resolve the degeneracy 
of vacua and stabilize the axion ${\rm Im}(U)$.
The SUSY vacuum or SUSY-breaking vacuum is selected depending on parameters in the potential.
Low-scale SUSY breaking is also possible when additional proper nonperturbative 
effects are involved.

We have also studied the model with the K\"ahler modulus $T$.
This model has the flat direction along both ${\rm Re}(T)$ and ${\rm Im}(T)$ at the leading level.
The SUSY vacuum and SUSY-breaking vacuum are degenerate, but 
the SUSY vacuum corresponds to the decompactification limit ${\rm Re}(T) \rightarrow \infty$.
In this model, the modulus $F$ term depends only on ${\rm Re}(T)$.
The real part ${\rm Re}(T)$ can be stabilized by inclusion of $\alpha'$ corrections 
and loop effects due to ${\rm Re}(T)$-dependent gaugino and sfermion masses.
However, the axion ${\rm Im}(T)$ remains massless at this stage.

We can extend the model with the single K\"ahler modulus to 
the models with many K\"ahler moduli.
Their real parts can be stabilized in a similar mechanism, 
but many axionic parts would remain light.
Such axions would be interesting, e.g., for candidates of 
dark matter and the QCD axion.
Also, one of the light axions could derive the cosmological inflation
if a proper potential is generated.
Moreover, these axions would be interesting from the viewpoint of a
string axiverse.
Such axion phenomenology would be studied elsewhere.

\section*{Acknowledgments}
T.K. was supported in part by the Grant-in-Aid for Scientific Research 
 No.~26247042  and No.~17H05395 from the Ministry of Education, Culture, Sports,
 Science and Technology in Japan. 
H.O. was supported in part by a Grant-in-Aid for Young Scientists (B) (No.~17K14303) 
from Japan Society for the Promotion of Science.

\appendix
\section{Moduli stabilization in the KKLT scenario}
\label{app}
In this Appendix, we present the stabilization of complex structure moduli and K\"ahler 
moduli in the setup of Sec.~\ref{sec:2}. 
To see the typical mass scale of complex structure moduli and K\"ahler moduli, 
we consider one of the complex structure moduli $U$ and overall K\"ahler modulus $T$. 
In  $4$D ${\cal N}=1$ effective supergravity, 
those are stabilized by the nonperturbative superpotential of the form
\begin{align}
W=B_0+B_1e^{-b_1T} U +B_2e^{-b_2T},
\end{align}
and the K\"ahler potential is given by
\begin{align}
K=-3\ln (U+\bar{U})-3\ln (T+\bar{T}).
\end{align}
Here, $C_1$ in Eq.~(\ref{eq:KW1}) corresponds to {$B_1e^{-b_1 \langle T \rangle}$}, and 
we assume that $B_0$ is a real constant for simplicity, 
although it is a function of $T$, in general.\footnote{When $B_0$ is a function of $T$, the overall K\"ahler modulus 
could be stabilized at the racetrack minimum.} 
Then, the total scalar potential becomes
\begin{align}
V=e^{K}\left(K^{U\bar{U}}|D_UW|^2+K^{T\bar{T}}|D_TW|^2-3|W|^2\right),
\end{align}
where $D_I=\partial_I +\partial_I K$, $I=T,U$, are the K\"ahler covariant derivatives and 
$K^{T\bar{T}}$ is the inverse of K\"ahler metric $K_{T\bar{T}}$. 

We first stabilize the overall K\"ahler modulus by the superpotential
\begin{align}
W^{(T)}=B_0+B_2e^{-b_2T},
\end{align}
which is the same as the KKLT scenario. 
It can be stabilized at the minimum satisfying
\begin{align}
D_TW=-b_2 B_2e^{-b_2T} -\frac{3}{T+\bar{T}}W^{(T)}=0,
\end{align}
which leads to the stabilization of $T$ as
\begin{align}
\langle T\rangle \simeq \frac{1}{b_2}\ln \left(\frac{2}{3B_0}\right),
\end{align}
in the limit of $b_2 \langle{\rm Re}(T)\rangle \gg 1$. 
At this minimum, the mass squared of canonically normalized $T$ is evaluated as
\begin{align}
m_{{\rm Re}(T)}^2=m_{{\rm Im}(T)}^2\simeq \frac{e^{\langle K\rangle}}{2} (K^{T\bar{T}})^2 |\partial_T\partial_T W|^2
\simeq 2e^{\langle K\rangle} (b_2\langle{\rm Re}(T)\rangle)^2 \langle W^{(T)}\rangle^2.
\end{align}
{For example, when $b_2 = 2 \pi$ we obtain $m_{{\rm Re}(T)}^2= 2e^{\langle K\rangle} (2\pi\langle{\rm Re}(T)\rangle)^2 \langle W^{(T)}\rangle^2$.}

Let us next stabilize $U$ at the fixed minimum of $T$. 
In this case, by setting $C_0=B_0+B_2e^{-b_2\langle T\rangle}$ and $C_1=B_1e^{-b_1\langle T\rangle}$, 
the superpotential is the same as in Eq.~(\ref{eq:KW1}) except for the factor $(T+\bar{T})^{-3}$. 
It turns out that the mass squared of $U$ is given by
\begin{align}
m_{{\rm Re}(U)}^2\simeq 16e^{\langle K\rangle} (C_0)^2\simeq 16e^{\langle K\rangle} \langle W^{(T)}\rangle^2, 
\end{align}
which is much lighter than $T$ in the limit of $b_2 \langle{\rm Re}(T)\rangle \gg 3$.  
To obtain the correct vacua of both the moduli fields, 
we have to stabilize them simultaneously. 
However, our analysis is justified because of the mass hierarchy between them.

%\bibliography{ref}

\end{document}